\documentclass[aps,prd,10pt,superscriptaddress]{revtex4-1}
\usepackage{floatpag}
\usepackage{amsmath}
\usepackage{amsbsy}
\usepackage{times}
\usepackage{bm}
\usepackage{graphicx}

\begin{document}

\title{\huge Trapping of charged particles by Bessel beams}
\author{\Large Iwo Bialynicki-Birula}\email{\large birula@cft.edu.pl}
\affiliation{\large Center for Theoretical Physics, Polish Academy of Sciences, Al. Lotnik\'ow 32/46, 02-668 Warsaw, Poland}
\author{\Large Zofia Bialynicka-Birula}
\affiliation{\large Institute of Physics, Polish Academy of Sciences, Al. Lotnik\'ow 32/46, 02-668 Warsaw, Poland}
\author{\Large Nadbor Drozd}
\affiliation{\large Center for Theoretical Physics, Polish Academy of Sciences, Al. Lotnik\'ow 32/46, 02-668 Warsaw, Poland}

\maketitle

\section{Introduction}

There exist two well established methods to trap charged particles: the Penning trap \cite{deh} and the Paul trap \cite{paul}. In the Penning trap the particle is confined in space by a combination of static magnetic and electric fields. In the Paul trap the trapping is caused by a high frequency electric quadrupole field.
The subject of this article is to present a third mechanism for trapping charged particles --- trapping by beams of electromagnetic radiation. It was discovered some time ago \cite{bb,bbch,br,mhj} that a properly prepared beam acts as a "waveguide" for particles, confining their motion in the transverse directions. Similar phenomena occur in RF fields \cite{ger}. In all these cases, the essential role is played by the electric field configuration in the plane perpendicular to the beam axis (for nonrelativistic electrons, the magnetic field is less important). Particles are confined to the vicinity of the minimum-energy points. In particular, for beams of electromagnetic radiation carrying orbital angular momentum such points lie on the beam axis. One beam may confine particles only in the transverse direction. Two or three crossing beams may fully confine particles, acting as a substitute for the Paul trap.

Trapping of charged particles by beams of electromagnetic radiation is based on the same general principles as the Paul trap. In the Paul trap the quadrupole radio-frequency field is produced by a system of electrodes. In our case, the field is that of freely propagating electromagnetic beams. In both cases the essential role is played by two factors: the proper shape of the force field and fast oscillation or rotation. These features are clearly seen when we invoke the notion of the ponderomotive potential. This notion is applicable when the the motion of a charged particle can be unambiguously split into a slow drift and a fast quiver motion. Such a splitting is possible when a nonrelativistic particle is moving in a rapidly oscillating field. We shall restrict ourselves in this paper to monochromatic beams since they are especially well suited for trapping particles. The best example of such beams are Bessel beams which will be analyzed in detail.

\section{Electric and magnetic fields of Bessel beams}

Every monochromatic electromagnetic field can be represented by the following general expressions:
\begin{subequations}
\begin{align}\label{elmag1}
{\bm E}({\bm r},t)={\bm E}_c({\bm r})\cos(\omega t)+{\bm E}_s({\bm r})\sin(\omega t),\\
{\bm B}({\bm r},t)={\bm B}_c({\bm r})\cos(\omega t)+{\bm B}_s({\bm r})\sin(\omega t).
\end{align}
\end{subequations}
According to Maxwell equations the electric and magnetic field vectors are interdependent,
\begin{align}\label{elmag2}
\omega{\bm B}_c({\bm r})={\bm\nabla}\times{\bm E}_s({\bm r}),\quad \omega{\bm B}_s({\bm r})=-{\bm\nabla}\times{\bm E}_c({\bm r}).
\end{align}
Generically, the electric and magnetic field vectors at each point trace two ellipses spanned by the pairs of vectors $\{{\bm E}_c,{\bm E}_s\}$ and $\{{\bm B}_c,{\bm B}_s\}$.

Bessel beams are most easily analyzed with the use of the Riemann-Silberstein (RS) vector \cite{pwf}, defined as a complex combination of the electric and magnetic fields,
\begin{align}\label{rs}
{\bm F}({\bm r},t)={\bm E}({\bm r},t)+ic{\bm B}({\bm r},t).
\end{align}
This vector obeys the following equations:
\begin{align}\label{me}
\partial_t{\bm F}({\bm r},t)=c{\bm\nabla}\times{\bm F}({\bm r},t),\quad
{\bm\nabla}\!\cdot\!{\bm F}({\bm r},t)=0.
\end{align}
For monochromatic fields we can write
\begin{align}\label{2hel}
{\bm F}({\bm r},t)=e^{-i\omega t}{\bm F}_+({\bm r})+e^{i\omega t}{\bm F}_-({\bm r}),
\end{align}
and the Maxwell equations reduce to the two equations for Trkalian fields \cite{trkal}
\begin{align}\label{trkal}
c{\bm\nabla}\times{\bm F}_\pm({\bm r})=\pm\omega{\bm F}_\pm({\bm r}).
\end{align}
The two parts in (\ref{2hel}), as we show below, correspond to two helicities. By comparing the formulas (\ref{elmag1}) and (\ref{2hel}), we find
\begin{align}\label{rel}
{\bm E}_c&={\rm Re}{\bm F}_++{\rm Re}{\bm F}_-,\quad {\bm E}_s={\rm Im}{\bm F}_+-{\rm Im}{\bm F}_-,\\
{\bm B}_c&={\rm Re}{\bm F}_++{\rm Im}{\bm F}_-,\quad {\bm B}_s=-{\rm Im}{\bm F}_++{\rm Re}{\bm F}_-.
\end{align}
It follows from these formulas that for fields of definite helicity the ellipses traced by the electric and magnetic field vector coincide.

Following Ref.~\cite{beams} (with some small changes in notation), we shall describe the electromagnetic field of the Bessel beam in cylindrical coordinates, with the $z$ axis in the direction of propagation, in terms of the RS vector in the following form:
\begin{align}\label{bess}
&\left(\begin{array}{c}F_x(\rho,\phi,z,t)\\F_y(\rho,\phi,z,t)\\F_z(\rho,\phi,z,t)
\end{array}\right)
=E_0e^{-i\chi\omega t}\,e^{ik_\parallel z}\nonumber\\
&\times\left(\begin{array}{c}\kappa_-e^{i(M+1)\phi}J_{M+1}(k_\perp\rho)+
\kappa_+e^{i(M-1)\phi}J_{M-1}(k_\perp\rho)\\
-i\kappa_-e^{i(M+1)\phi}J_{M+1}(k_\perp\rho)+
i\kappa_+e^{i(M-1)\phi}J_{M-1}(k_\perp\rho)\\
-2ie^{iM\phi}J_M(k_\perp\rho)
\end{array}\right).
\end{align}
where
\begin{align}\label{def}
\kappa_\pm =\frac{\chi\sqrt{k_\parallel^2+k_\perp^2}\pm k_\parallel}{k_\perp}.
\end{align}
We shall consider only positive values of $M$ since Bessel beams with negative values can be obtained (up to an overall sign) by the following transformations of spacetime variables and helicity:
\begin{subequations}
\begin{align}\label{equiv}
 M\rightarrow-M,\quad\chi\rightarrow-\chi\quad\Longrightarrow\quad\phi\rightarrow -\phi,\quad t\rightarrow -t.
\end{align}
\end{subequations}
The formulation in terms of the RS vector enables one to introduce, by analogy with quantum mechanics, the classification of the solutions of Maxwell equations with the use of quantum numbers \cite{beams}. One may view these quantum numbers as characteristics of the photons making the beam. In the case of Bessel beams these quantum numbers are the eigenvalues of the following four operators (we dropped $\hbar$ in the definitions): the component of momentum in the direction of propagation $\hat{k}_\parallel$, the length of momentum in the transverse direction $\hat{k}_\perp$, the projection of the total angular momentum (orbital plus spin) on the direction of propagation $\hat{M}_z$, and the helicity operator (the projection of the total angular momentum on the direction of momentum) $\hat\chi$,
\begin{align}
\hat{k}_\parallel &=-i\partial_z,\quad \hat{k}_\perp=\sqrt{-\partial_x^2-\partial_y^2},\label{op1}\\
\nonumber\\
\hat{M}_z&=-i(x\partial_y-y\partial_x)+\left[\!\begin{array}{ccc}
0 & -i & 0\\
i & 0 & 0\\
0 & 0 & 0\end{array}
\right],\label{op2}\\
\nonumber\\
{\hat\chi}&=\frac{1}{2\pi^2}\int\!d^3r'\frac{1}{|\bm r-\bm r'|^2}\left[\!\begin{array}{ccc}
0 & -\partial_{z'} & \partial_{y'}\\
\partial_{z'} & 0 & -\partial_{x'}\\
-\partial_{y'} & \partial_{x'} & 0\end{array}
\right].\label{op3}
\end{align}
The $3\times 3$ matrix in Eq.~(\ref{op2}) represents the $z$ component of the photon spin in the vector representation. The representation of the helicity operator in the form (\ref{op3}) was given in \cite{phloc}. The factors $\exp[i(M\pm1)\phi]$ and $\exp[iM\phi]$ in the definition of Bessel beams reflect the decomposition of the total angular momentum $M$ into the components with three different orbital angular momenta.

\begin{figure}
\centering
\includegraphics[scale=1]{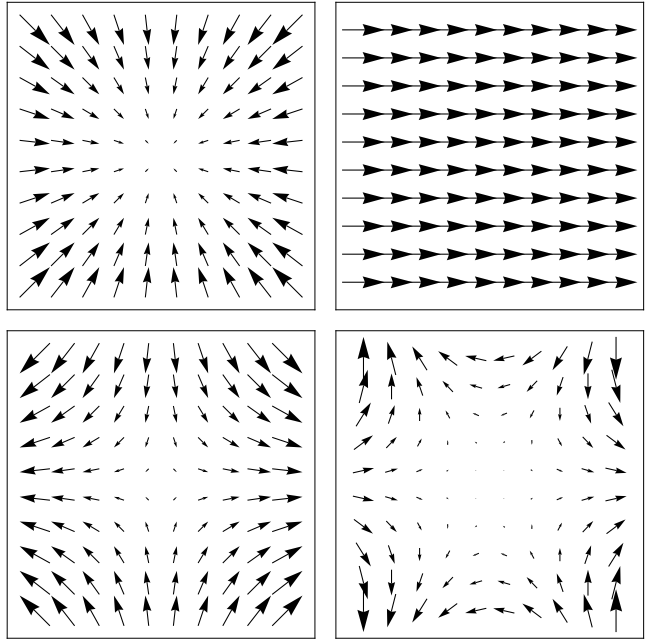}
\caption{Electric field vectors in the $xy$ plane near the beam axis for $\chi=1$ and for $M=0, 1, 2, 3$. This field configuration corresponds to $z=0$ and $t=0$. For other values the vectors would be rotated by the angle $(k_\parallel z-\chi\omega t)$. The top configurations correspond to $M=0$ and $M=1$. The bottom configurations correspond to $M=2$ and $M=3$.}\label{fig1}
\end{figure}

\begin{figure}
\centering
\includegraphics[scale=1]{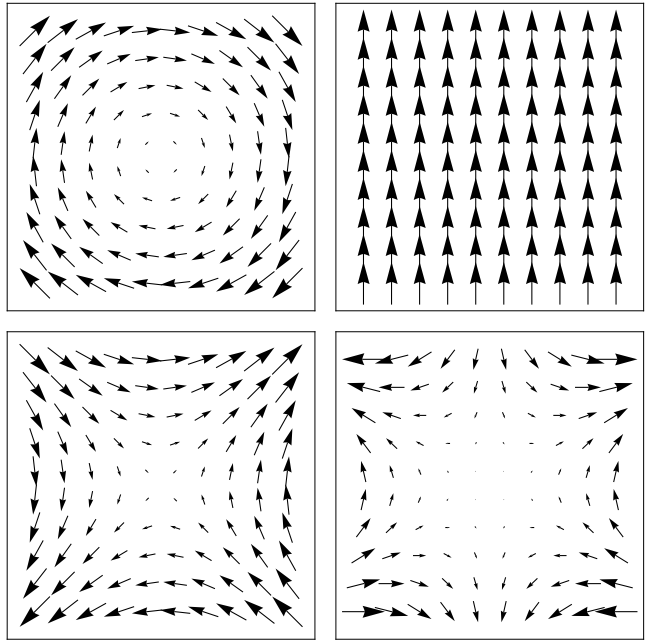}
\caption{Electric field vectors in the $xy$ plane as in Fig.~\ref{fig1} but a quarter of the cycle later. Each field vector is rotated counterclockwise by $\pi/2$.}\label{fig2}
\end{figure}

\begin{figure}
\centering
\includegraphics[scale=1]{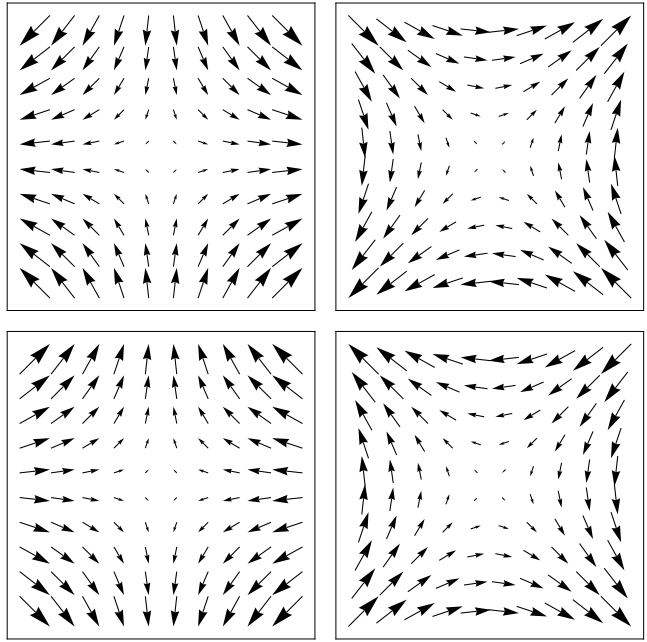}
\caption{Snapshots of the electric field vector in the transverse plane for $M=2$ and $\chi=1$, taken at $t=0,\,T/4,\,T/2,\,3T/4$, where $T$ is the period of the wave.}\label{fig3}
\end{figure}

In classical physics quantum numbers are just convenient labels attached to the solutions of Maxwell equations. Thus, each Bessel beam (\ref{bess}) is characterized by the following four parameters: a real parameter $k_\parallel$, a positive parameter $k_\perp$, an integer $M$, and $\chi$ ($\pm 1$). Every Bessel beam is monochromatic with the frequency $\omega=c\sqrt{k_\parallel^2+k_\perp^2}$. The amplitude $E_0$ in Eq.~(\ref{bess}) measures the strength of the beam in volt per meter (1 V/m corresponds to the beam intensity 2.65$\times10^{-3}\,{\rm W/m}^2$). The dimensionless electric and magnetic field vectors for the Bessel beam are obtained by taking the real and imaginary parts in Eq.~(\ref{bess}) and dropping the factor $E_0$,
\begin{subequations}\label{bessem}
\begin{align}
\left(\begin{array}{c}{\cal E}_x\\{\cal E}_y\\{\cal E}_z\end{array}\right)
&=\left(\begin{array}{c}\kappa_- c_{M+1}+\kappa_+ c_{M-1}\\
\kappa_- s_{M+1}-\kappa_+ s_{M-1}\\
2s_{M}\end{array}\right),\\
\left(\begin{array}{c}{\cal B}_x\\{\cal B}_y\\{\cal B}_z\end{array}\right)
&=\left(\begin{array}{c}\kappa_- s_{M+1}+\kappa_+ s_{M-1}\\
-\kappa_- c_{M+1}+\kappa_+ c_{M-1}\\
-2c_{M}\end{array}\right),
\end{align}
\end{subequations}
where
\begin{subequations}
\begin{align}\label{defem}
c_M &=\cos(k_\parallel z-\chi\omega t+M\phi)J_M\left(k_\perp\rho\right),\\
s_M &=\sin(k_\parallel z-\chi\omega t+M\phi)J_M\left(k_\perp\rho\right).
\end{align}
\end{subequations}
In Fig.~\ref{fig1} we show the electric field configuration in the transverse plane near the beam axis for different values of $M$. Note that for $M=2$ the configuration of electric forces is the same as of the gravitational forces on the saddle surface in the mechanical model of the Paul trap \cite{paul}. Therefore, at least in this case we can expect trapping of particles in the transverse plane. To exhibit rotation of field vectors in time, we plotted these vectors in Fig.~\ref{fig2} at the time one quarter of the cycle later. In Fig.~\ref{fig3} we show the rotation of the field vectors for $M=2$.

\section{Motion of charged particles in a Bessel beam}

All field components of the Bessel beam with $|M|>1$ vanish along the $z$ axis since in these cases all Bessel functions in Eq.~(\ref{bess}) vanish at the origin. The motion of charged particles in Bessel beams in the vicinity of this line was thoroughly studied in Ref.~\cite{bbch} by solving numerically the standard equations of motion
\begin{align}\label{eqm}
m\frac{d^2x^\mu(s)}{ds^2}=ef^{\mu\nu}(x(s))\frac{dx_\nu(s)}{ds},
\end{align}
where $s$ is the proper time. These equations can be rewritten in the following dimensionless form
\begin{align}\label{eqmd}
\frac{d^2\xi^\mu(\tau)}{d\tau^2}=\gamma\, \varphi^{\mu\nu}(\xi(\tau))\frac{d\xi_\nu(\tau)}{d\tau},
\end{align}
where the particle coordinates $\xi^\mu$ are measured in units of the longitudinal wave length $2\pi/k_\parallel$ and the dimensionless proper time $\tau$ is measured in units of the inverse circular frequency $\tau=\omega s$. The dimensionless effective coupling constant $\gamma$ is
\begin{align}\label{gamma}
\gamma=\frac{eE_0k_\parallel}{m\omega^2}
=\frac{eE_0/k_\parallel}{mc^2(1+(k_\perp/k_\parallel)^2)}.
\end{align}
The dimensionless electromagnetic field tensor $\varphi^{\mu\nu}$ is composed of the dimensionless electric and magnetic components given by Eqs.~(\ref{bessem}). The parameter $\gamma$ is usually very small. It is approximately equal to the ratio of the energy gain (i.e. the energy that the particle acquires moving under the influence of the electric field $E_0$ along the distance of $1/k_\parallel$) to the particle rest energy $mc^2$. The particle trajectories scaled by the wavelength with initial velocities scaled by the speed of light will look the same for different situations provided the value of $\gamma$ remains the same. The smaller the mass, the more effective is the action of the electromagnetic wave. Therefore, in all our specific examples we shall consider electrons.
In what follows we shall consider nonrelativistic electrons whose motion is governed mainly by the electric field. One beam may confine the motion of an electron only in the transverse directions --- the electron is guided along the direction of propagation. For full trapping we will need at least two beams.

\begin{figure}
\centering
\includegraphics[scale=0.45]{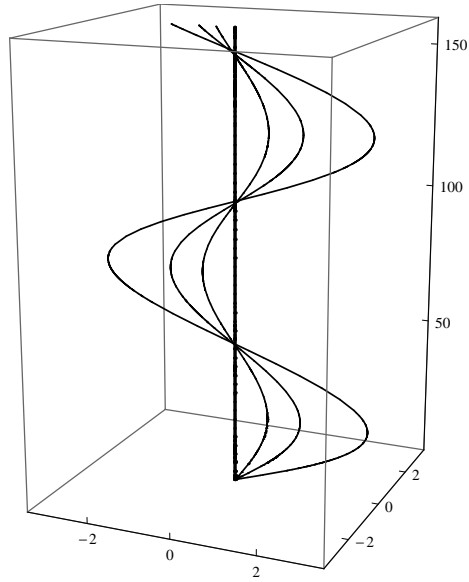}
\caption{Three trajectories of an electron guided by the Bessel beam with $M=2$ and $\chi=1$ obtained for different initial transverse velocities. All distances are measured in units of the longitudinal wavelength. Thick line represents the beam axis. Note a large difference in the scaling of transverse and longitudinal distances.}\label{fig4}
\end{figure}

Typical trajectories of an electron guided by the Bessel beam are shown in Fig.~\ref{fig4}. For Bessel beams describing the propagation of light, the effective transverse dimension $2\pi/k_\perp$ is much larger than the longitudinal wave length. This means that the dimensionless parameter $k_\perp/k_\parallel$ is usually very small. The trajectories shown in Fig.~\ref{fig4} are obtained for $k_\perp/k_\parallel=0.01$ and $\gamma=0.00001$. For visible light, this value of $\gamma$ corresponds to the beam intensity of the order of $10^{13}{\rm W/m}^2$. This high intensity resulted from our choice of rather high transverse velocities that were taken for illustrative purposes. The three trajectories guided by the beam start at the same point near the beam axis and with the same initial longitudinal velocity $\dot{z}(0)=0.001c$ but they have different initial transverse velocities $\dot{x}(0)=0.0002c,\;0.0001c$, and $0.00005c$. Trapping of electrons with initial transverse velocity exceeding the value $0.001c$  will require the increase of the beam intensity. For moderate beam intensities, relativistic electrons will not be trapped.
\begin{figure}
\centering
\includegraphics[scale=0.5]{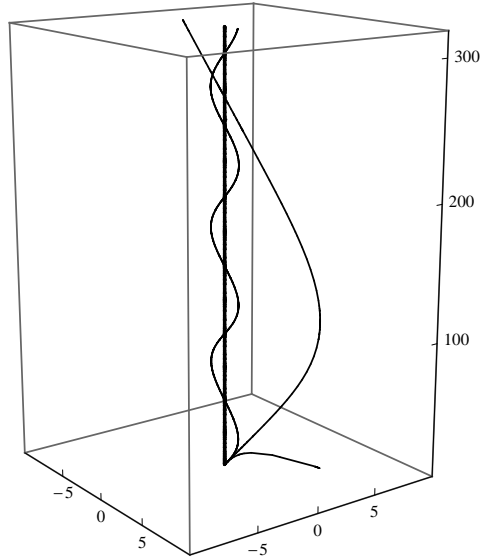}
\caption{Four trajectories of an electron guided by the Bessel beams with $M=0, 1, 2$ and $3$. The tightly trapped trajectories for $M=0$ and $M=2$ are indistinguishable. Trapping is less pronounced for $M=3$ (wide trajectory). There is no trapping for $M=1$ (runaway trajectory).}\label{fig5}
\end{figure}
\begin{figure}
\centering
\includegraphics[scale=0.3]{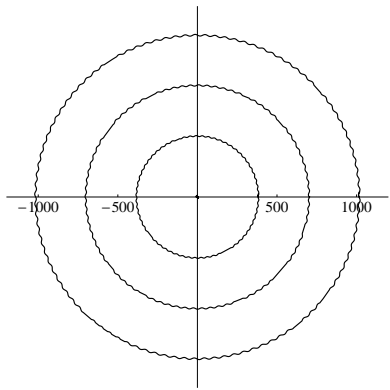}
\caption{Projections on the transverse plane of the three trajectories of an electron guided by the Bessel beams with $M=2$. The trajectories started with the same initial velocities in the vicinity of the zeros of the Bessel function.}\label{fig6}
\end{figure}

In Fig.~\ref{fig5} we plotted the trajectories trapped near the beam center for different $M$ obtained when $\dot{x}(0)=0.0001c$ and with the values of the remaining parameters the same as in Fig.~\ref{fig4}. We can easily explain qualitatively these results, with the help of the formulas (\ref{bessem}) and Fig.~\ref{fig1}. First of all, when $k_\perp/k_\parallel$ is small, only the Bessel function $J_{M-\chi}$ contributes significantly to the transverse components of the fields. This observation underscores the role of the {\em orbital angular momentum}. For realistic Bessel beams only one orbital angular momentum component, with $M-\chi$,  plays the dominant role. Therefore, for $\chi=1$, the trapping will occur only for $M=0,\;2$ and 3 but not for $M=1$. This is so because in the last case the dominant part comes from the Bessel function $J_0$ and near the beam center the field configuration is that of a circularly polarized plane wave. Trapping is the best for $M=0$ and $M=2$ and for $\chi=1$ because only in these two cases the dominant role in the formula (\ref{bess}) is played by the Bessel function $J_1$. Therefore, the field strength increases linearly with the distance from the beam axis, as in the Paul's mechanical model. For $M=3$ the trapping is weak, because the Bessel function $J_2$ increases more slowly near the beam center, so that the radius of the effective potential well is larger. With the increase of $M$, the trapping will become weaker and weaker.

The problem of an electron motion near the Bessel beam axis was solved explicitly in Ref.~\cite{bb} in classical and in quantum mechanics for the fields described in the paraxial approximation. The analytic solution in this case was possible due to the fact that the field configuration close to the center of the beam with $M=2$ and $\chi=1$ is particularly simple---the field increases linearly with the distance from the beam axis. The main feature of the analytic solutions is the separation of the longitudinal motion from the motion in the transverse plane. The motion in the transverse plane is bounded whereas the longitudinal motion decomposes into oscillations and a uniform motion with velocity determined by the initial conditions. The analytic solutions of the equations of motion (\ref{eqm}) in this simplified model fully confirm our numerical results.

Bessel functions vanish not only at the center of the beam but they also have a succession of other zeros. It turns out that charged particles may be trapped in regions corresponding to these zeroes. In Fig.~\ref{fig6} we show the four trajectories trapped at the first four zeroes for $M=2$ and $\chi=1$.

\section{Ponderomotive potential}

The trajectories of particles trapped by Bessel beams in the nonrelativistic regime can be reproduced remarkably well in the ponderomotive approximation. The notion of the ponderomotive potential introduced by Gaponov and Miller \cite{gm} and independently by Boot and Harvie \cite{bh} is very useful when the particle moves slowly in an inhomogeneous, rapidly oscillating electromagnetic field. The approximation is based on the decomposition of the particle trajectory into a sum ${\bm R}(t)+{\bm r}(t)$, where the first part is slowly varying and the second part describes small but fast oscillations with zero mean value. We start from the equations of motion with the Lorentz force,
\begin{align}\label{pma0}
{\ddot{\bm R}}+{\ddot{\bm r}}=\frac{e}{m}\left[{\bm E}({\bm R}+{\bm r},t)+\left({\dot{\bm R}}+{\dot{\bm r}}\right)\times{\bm B}({\bm R}+{\bm r},t)\right].
\end{align}
The necessary condition for the applicability of the ponderomotive approximation is $|{\bm v}\!\cdot\!{\bm\nabla}f|/|\omega f|\ll 1$, where $f$ is the force field. Under this assumption, one may expand the fields in the equations of motion around the slow trajectory ${\bm R}(t)$ keeping only linear terms in ${\bm r}(t)$. In the lowest order, using the notation of Eqs.~(\ref{elmag1}) and (\ref{elmag2}), we obtain
\begin{align}\label{pma}
{\ddot{\bm r}}=\frac{e}{m}\left[{\bm E}_c({\bm R})\cos(\omega t)+{\bm E}_s({\bm R})\sin(\omega t)\right].
\end{align}
After disregarding the slow variation of ${\bm R}$ with time, we obtain by a straightforward integration
\begin{align}\label{pma1}
{\bm r}=-\frac{e}{m\omega^2}\left[{\bm E}_c({\bm R})\cos(\omega t)+{\bm E}_s({\bm R})\sin(\omega t)\right],
\end{align}
where we dropped the integration constants to obtain zero mean values. In the next order, the equation reads
\begin{align}\label{pma2}
{\ddot{\bm R}}&=\frac{e}{m}\Big\{\Big[\left({\bm r}\cdot{\bm\nabla}\right){\bm E}_c({\bm R})+\frac{1}{\omega}{\dot{\bm r}}\times\left({\bm\nabla}\times{\bm E}_s({\bm R})\right)\Big]\cos(\omega t)\nonumber\\
&+\Big[\left({\bm r}\cdot{\bm\nabla}\right){\bm E}_s({\bm R})-\frac{1}{\omega}{\dot{\bm r}}\times\left({\bm\nabla}\times{\bm E}_c({\bm R})\right)\Big]\sin(\omega t)\Big\}.
\end{align}
In the final step, after inserting the solutions of Eq.~(\ref{pma}) for ${\bm r}$ and ${\dot{\bm r}}$, we perform time averaging of the right hand side over one period of fast oscillations $2\pi/\omega$ and we obtain
\begin{align}\label{pma3}
{\ddot{\bm R}}&=-\frac{e^2}{2m^2\omega^2}\Big[\left({\bm E}_c({\bm R})\cdot{\bm\nabla}\right){\bm E}_c({\bm R})+{\bm E}_s({\bm R})\times\left({\bm\nabla}\times{\bm E}_s({\bm R})\right)\nonumber\\
&+\left({\bm E}_s({\bm R})\cdot{\bm\nabla}\right){\bm E}_s({\bm R})+{\bm E}_c({\bm R})\times\left({\bm\nabla}\times{\bm E}_c({\bm R})\right)\Big]\nonumber\\
&=-\frac{e^2}{4m^2\omega^2}{\bm\nabla}\left[{\bm E}_c({\bm R})^2+{\bm E}_s({\bm R})^2\right]=-{\bm\nabla}V_p({\bm R}),
\end{align}
where we used of the vector identity
\begin{align}\label{id}
{\bm E}\times\left({\bm\nabla}\times{\bm E}\right)=\frac{1}{2}{\bm\nabla}{\bm E}^2-\left({\bm E}\cdot{\bm\nabla}\right){\bm E}.
\end{align}
The ponderomotive potential $V_p({\bm R})$ can also be expressed as the square of the electric field averaged over time
\begin{align}\label{pf}
V_p({\bm R}) = \frac{e^2}{2m\omega^2}\langle{\bm E}({\bm R},t)\cdot{\bm E}({\bm R},t)\rangle.
\end{align}
Despite the fact that only the electric field enters into this formula, the ponderomotive potential contains also a contribution from the magnetic field, as seen from our derivation. In our dimensionless variables the equations of motion have the form
\begin{align}\label{pfeq}
{\ddot{\bm\xi}}=-\frac{\gamma^2}{2}{\bm\nabla}\langle{\bm{\mathcal E}}({\bm\xi},\tau)\cdot{\bm{\mathcal E}}({\bm\xi},\tau)\rangle.
\end{align}

The use of ponderomotive approximation requires some care in handling  the initial conditions. There is no need to adjust the initial value of the slowly varying part of the trajectory (the term ${\bm R}$ in Eq.~(\ref{pma})). We can always identify ${\bm R}(t=0)$ with the initial position of the exact trajectory because the amplitude of fast oscillations is small. However, the adjustment of the initial velocity is necessary because the velocity of fast oscillations is large and it gives a significant contribution to the true velocity. This adjustment is especially important when the trajectory starts in a region where the field is strong. According to Eq.~(\ref{pma1}), ${\dot{\bm r}}$ can be related to the electric field by the following formula
\begin{align}\label{v}
{\dot{\bm r}}&=\frac{e}{m\omega^2}\left[{\bm E}_c({\bm R})\sin(\omega t)-{\bm E}_s({\bm R})\cos(\omega t)\right]\nonumber\\
&=-\frac{e}{m\omega^2}\left[{\bm E}_c({\bm R})\cos(\omega t+\pi/2)+{\bm E}_s({\bm R})\sin(\omega t+\pi/2)\right]\nonumber\\
&=-\frac{e}{m\omega^2}{\bm E}({\bm R},t+\pi/2/\omega).
\end{align}
Since the ponderomotive approximation reproduces only the slowly changing part of the trajectory, the initial velocity of the slow motion ${\dot{\bm R}}$ must be identified with the difference of the exact solution and ${\dot{\bm r}}$. The significance of the initial velocity adjustment will be illustrated with the examples given in the next section.

\begin{figure}
\centering
\includegraphics[scale=0.5]{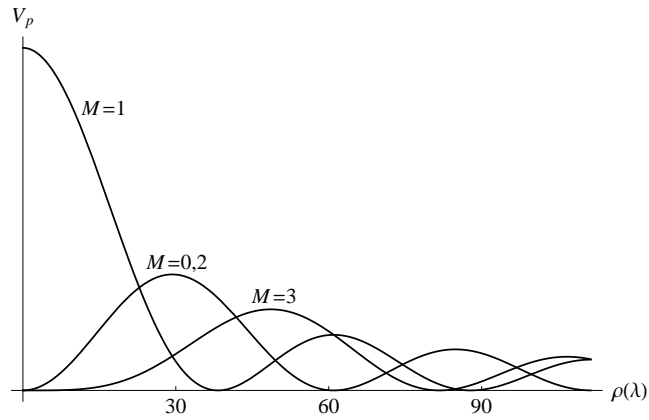}
\caption{Ponderomotive potential for the Bessel beams as a function of the distance from beam axis $\rho$.}\label{fig7}
\end{figure}

The ponderomotive potential for a Bessel beam is a function of the distance from the axis only,
\begin{align}\label{pfb}
V_p(\rho) = V_0\left[\left(\kappa_+ J_{M-1}(k_\perp\rho)\right)^2 + 2 \left(J_M(k_\perp\rho)\right)^2 +\left(\kappa_- J_{M+1}(k_\perp\rho)\right)^2\right],
\end{align}
where $V_0$ is the amplitude determined by the beam intensity; in our dimensionless variables, $V_0=\gamma^2/2$. For realistic Bessel beams, when $k_\perp\ll k_\parallel$, only one term in this formula contributes significantly. This term represents the contribution from the component with orbital angular momentum $M-\chi$. The ponderomotive potentials for $\chi=1,\;k_\perp/k_\parallel=0.01$, and for different values of $M$ are shown in Fig.~\ref{fig7}. The minima correspond to the zeros of the dominant Bessel function. Only the minima at $\rho=0$ for $M>1$ are exactly equal zero, the remaining ones are close to zero.

\begin{figure}
\centering
\includegraphics[scale=0.5]{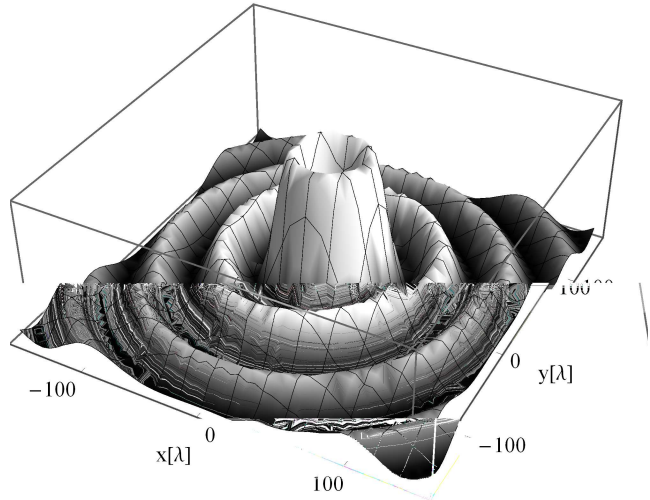}
\caption{Ponderomotive potential for the Bessel beam with $M=2$ and $\chi=1$ as a function of $x$ and $y$.}\label{fig8}
\end{figure}

The trajectories obtained by integration of exact equations, depicted in Fig.~\ref{fig4} and in Fig.~\ref{fig5} coincide with those obtained from the ponderomotive potential (\ref{pfb}). The vertical line in those figures is the line of the potential minima (except for $M=1$ where it is a maximum). For $M>1$ the beam axis is an obvious potential minimum, since the whole electromagnetic field vanishes there. For $M=0$ the $z$-component of the field does not vanish on the axis but its contribution is very small when $k_\perp/k_\parallel$ is small. The notion of the ponderomotive potential enables one to understand the existence of a limiting transverse particle velocity that allows for trapping. The transverse kinetic energy cannot exceed the height of the ponderomotive potential.

The ponderomotive potential for the Bessel beams, in addition to the potential well along the axis, has a succession of potential valleys corresponding to the minima in Fig.~\ref{fig7}. In Fig.~\ref{fig8} we show the potential for $M=2$ and $\chi=1$ as a function of $x$ and $y$ where the potential valleys are clearly seen. The trajectories depicted in Fig.~\ref{fig6} are trapped in those valleys and they are reproduced very well by solving the equations with the ponderomotive potential.

So far we considered trapping by beams carrying definite angular momentum. Such beams are most natural in this context since the ponderomotive potential has easily identified minima. However, as shown in the next section, trapping is found also for field configurations that do not have definite angular momentum.

\section{Trapping of particles by superpositions of Bessel beams}

\subsection{Trapping of particles along a helix}

\begin{figure}
\centering
\includegraphics[scale=0.6]{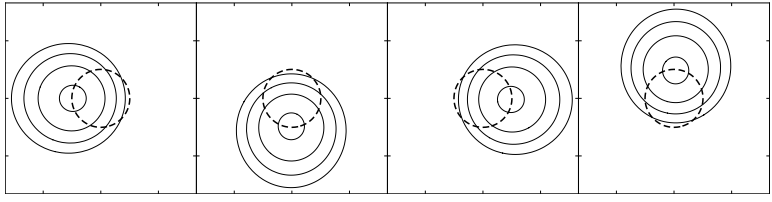}
\caption{Contour lines of the ponderomotive potential in the $xy$ plane created by a superposition of the Bessel beam with $M=2$ and a plane wave with the same frequency $\omega$ drawn for four values of $z$. The dashed circle represents the projection of the helix on the $xy$ plane.}\label{fig9}
\end{figure}

\begin{figure}
\centering
\includegraphics[scale=0.7]{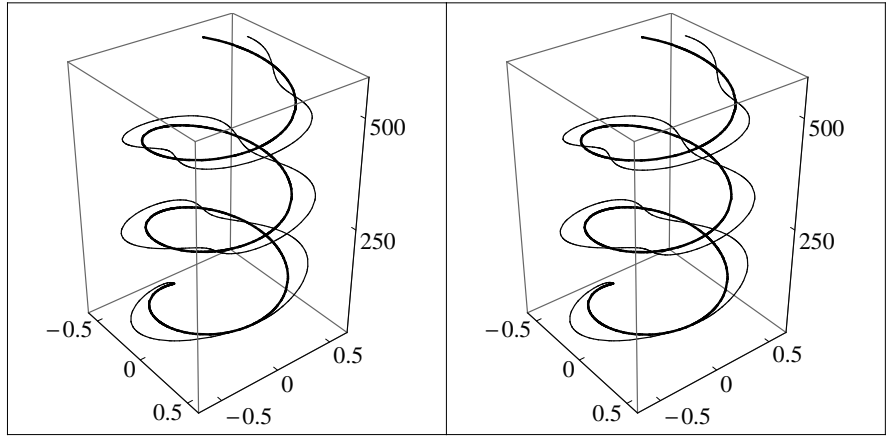}
\caption{Motion of a slow particle guided by the helix (thick line) created by a superposition of the Bessel beam and a plane wave. There are no noticeable differences between the exact solution (left) and the ponderomotive approximation (right).}\label{fig10}
\end{figure}

\begin{figure}
\centering
\includegraphics[scale=0.7]{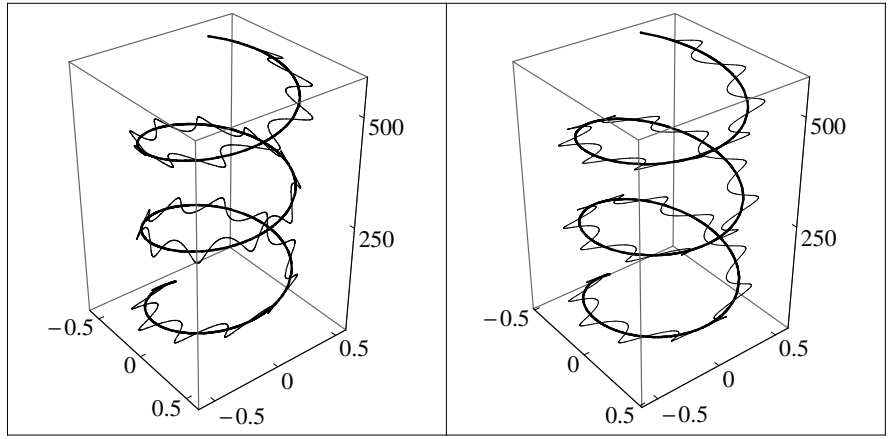}
\caption{Motion of a fast particle in the field as in Fig.~\ref{fig10}. There is a clear difference between the exact (left) and the approximate solution (right) but the ponderomotive potential still reproduces qualitatively the trapping near the helix.}\label{fig11}
\end{figure}

We shall now consider the motion of an electron in the superposition of a Bessel beam and a circularly polarized plane wave. We assume that both waves propagate in the same direction and have the same frequency. The dimensionless electric and magnetic fields are in this case
\begin{subequations}
\begin{align}\label{bessem1}
\left(\begin{array}{c}{\cal E}_x\\{\cal E}_y\\{\cal E}_z\end{array}\right)
&=\left(\begin{array}{c}a\cos(\omega(z/c-t)+(\kappa_- c_{M+1}+\kappa_+ c_{M-1})\\
-a\sin(\omega(z/c-t)+(\kappa_- s_{M+1}-\kappa_+ s_{M-1})\\
2s_{M}\end{array}\right),\\
\left(\begin{array}{c}{\cal B}_x\\{\cal B}_y\\{\cal B}_z\end{array}\right)
&=\left(\begin{array}{c}a\sin(\omega(z/c-t)+(\kappa_- s_{M+1}+\kappa_+ s_{M-1})\\
a\cos(\omega(z/c-t)+(-\kappa_- c_{M+1}+\kappa_+ c_{M-1})\\
-2c_{M}\end{array}\right),
\end{align}
\end{subequations}
where $a$ measures the relative strength of the plane wave as compared to the Bessel beam. The ponderomotive potential is given by the formula
\begin{align}\label{helpot}
&V_p(\rho,\phi,z) = V_0\big[\left(\kappa_+ J_{M-1}(k_\perp\rho)\right)^2
 +2 \left(J_M(k_\perp\rho)\right)^2
 +\left(\kappa_- J_{M+1}(k_\perp\rho)\right)^2\nonumber\\
&+a^2+2a\kappa_+J_{M-1}(k_\perp\rho)\cos\left((M-1)\phi-(\omega-k_\parallel)z\right)\big].
\end{align}
The minima of $V_p(\rho,\phi,z)$ can be easily found. The minimum in variables $\phi$ and $z$ is reached when the cosine attains the lowest value -1. The minimum in $\rho$ for this value of the cosine can be easily found numerically. These condition define the helix $z=(\omega-k_\parallel)(\phi+\pi)/(M-1)$. The line of the helix traces the minima of the ponderomotive potential and is the analog of the beam axis for a Bessel beam.  The helical motion of these minima along the beam axis is visualized in Fig.~\ref{fig9}. Therefore, in this case we can expect that the electrons will be guided along the helical line. However, this requires stronger fields. For electrons with initial transverse velocities the same as for a single Bessel beam we increased $\gamma$ to the value $0.0005$. In Fig.~\ref{fig10} we show the trajectories for $M=2$, $\chi=1$, and $a=3$ obtained by solving exact equations and those with the ponderomotive potential. The small parameter $k_\perp/k_\parallel$ determines the radius of the helix and to make this figure more transparent we have increased this parameter tenfold from the value used in previous figures.

In order to trap more energetic particles, we need stronger fields. As an example, we have chosen the initial transverse velocity as $0.01$c and we have increased the field strength to $\gamma=0.01$. In Fig.~\ref{fig11} we show that the particle is trapped around the helix. The ponderomotive approximation is still applicable but it does not reproduce exactly the fine details of the trajectories.

\subsection{Trapping of particles by crossed Bessel beams}

\begin{figure}
\centering
\includegraphics[scale=0.6]{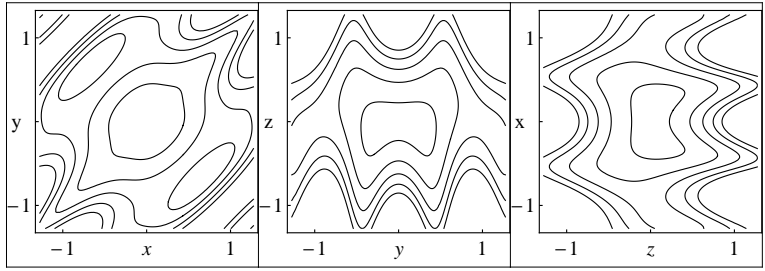}
\caption{Contour lines in the $xy,\,yz$, and $zx$ planes of the ponderomotive potential for two crossed Bessel beams with $M=0$ propagating in the $x$ and $y$ directions. There is a potential minimum at the crossing point.}\label{fig12}
\end{figure}

\begin{figure}
\centering
\includegraphics[scale=0.6]{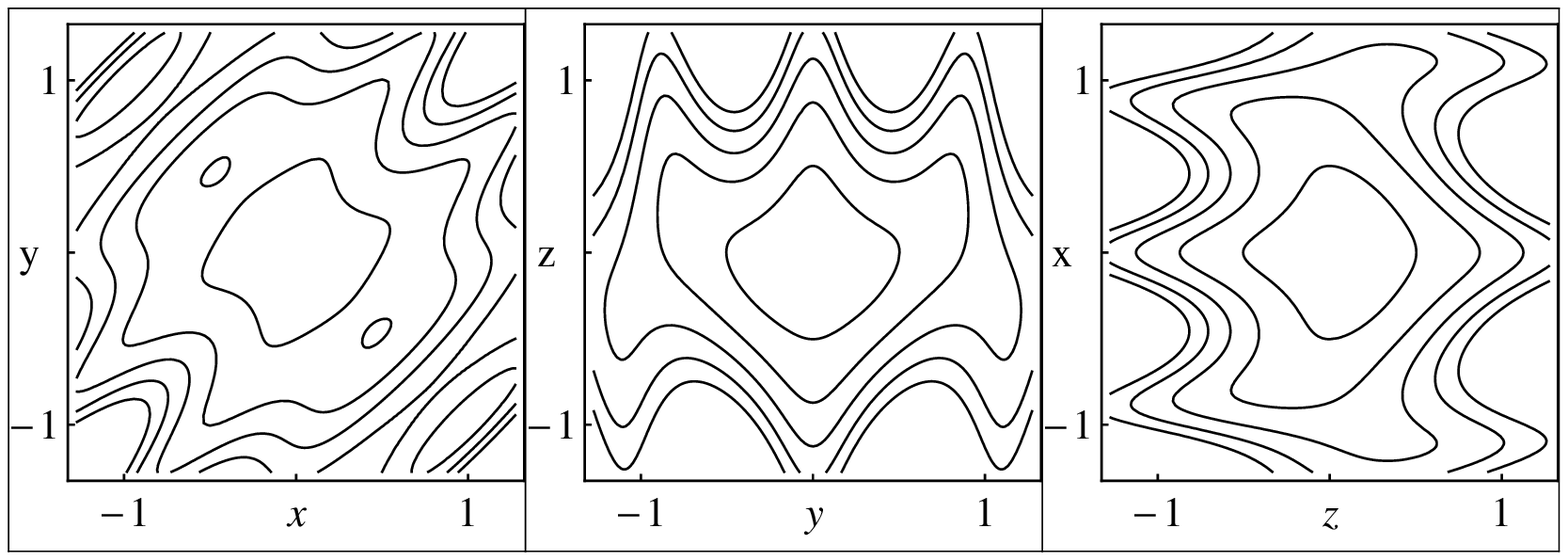}
\caption{Contour lines in the $xy,\,yz$, and $zx$ planes of the ponderomotive potential for two crossed Bessel beams with $M=2$ propagating in the $x$ and $y$ directions. There is a potential minimum at the crossing point.}\label{fig13}
\end{figure}

\begin{figure}
\centering
\includegraphics[scale=0.5]{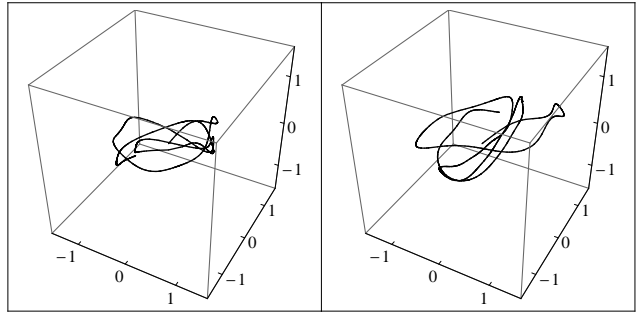}
\caption{Trajectories of a particle trapped in the vicinity of the potential minimum. Exact trajectory (left) is completely different from the trajectory obtained in the ponderomotive approximation without the velocity adjustment (right).}\label{fig14}
\end{figure}

\begin{figure}
\centering
\includegraphics[scale=0.5]{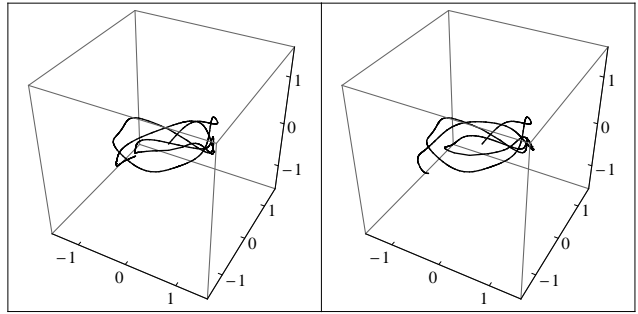}
\caption{The trajectory obtained in the ponderomotive approximation with the velocity adjustment (right) is much closer to the exact trajectory (left) than in Fig.~\ref{fig14}.}\label{fig15}
\end{figure}

In all previous cases the trapping occurred only in the transverse direction. The trapping in all directions can be achieved by using crossed Bessel beams. Crossed Bessel beams trap particles in the vicinity of the crossing point. We have chosen two identical Bessel beams propagating in the $x$ and $y$ direction. In this case, the ponderomotive potential is given by the formula:
\begin{align}\label{vpcross}
&V_p=V_0\big[\kappa_+^2\left(J_{M-1}^2(\rho_x)+J_{M-1}^2(\rho_y)\right)
+\kappa_-^2\left(J_{M+1}^2(\rho_x)+J_{M+1}^2(\rho_y)\right)\nonumber\\
&-2\kappa_+J_{M}(\rho_x)J_{M-1}(\rho_y)\cos(\psi+\phi_y)
+2\kappa_-J_{M}(\rho_x)J_{M+1}(\rho_y)\cos(\psi-\phi_y)\nonumber\\
&-2\kappa_+J_{M-1}(\rho_x)J_{M}(\rho_y)\sin(\psi-\phi_x)
-2\kappa_-J_{M+1}(\rho_x)J_{M}(\rho_y)\sin(\psi+\phi_x)\nonumber\\
&-\kappa_+^2J_{M-1}(\rho_x)J_{M-1}(\rho_y)\sin(\psi-\phi_x+\phi_y)\nonumber\\
&+\kappa_-^2J_{M+1}(\rho_x)J_{M+1}(\rho_y)\sin(\psi+\phi_x-\phi_y)\nonumber\\
&-J_{M-1}(\rho_x)J_{M+1}(\rho_y)\sin(\psi-\phi_x-\phi_y)\nonumber\\
&+J_{M+1}(\rho_x)J_{M-1}(\rho_y)\sin(\psi+\phi_x+\phi_y)\nonumber\\
&+2\left(J_{M}^2(\rho_x)+J_{M}^2(\rho_y)\right)\big],
\end{align}
where we used the following abbreviations:
\begin{align}\label{abb}
&\phi_x=\arctan(z/y),\quad\phi_y=\arctan(x/z),\quad
\psi=k_\parallel(x-y)+M(\phi_x-\phi_y)\nonumber\\
&\rho_x=k_\perp\sqrt{y^2+z^2},\quad\rho_y=k_\perp\sqrt{z^2+x^2}.
\end{align}
Again the value $M=1$ is exceptional---it does not lead to trapping. The shapes of the potential wells for $M=0$ and $M=2$ are very similar, as shown in Fig.~\ref{fig12} and Fig.~\ref{fig13}). These shapes clearly indicate that the potentials produced by crossed Bessel beams can trap particles in the vicinity of the beam crossing point. Solutions of the equations of motion fully confirm these predictions. For higher values of $M$ trapping is less effective since the wells become shallower.

For crossed Bessel beams the ponderomotive approximation is less successful than for a single Bessel beam. In Fig.~\ref{fig14} and Fig.~\ref{fig15} the exact trajectory for $M=0$ is compared with the approximate trajectories. Even though the adjustment of the initial velocity significantly improves the result, there remain substantial differences. The discrepancy is most likely due to a rather complicated shape of the potential well which may lead to a chaotic behavior---small changes produce large effects.

\subsection{Trapping of particles by a standing Bessel wave}

\begin{figure}
\centering
\includegraphics[scale=0.8]{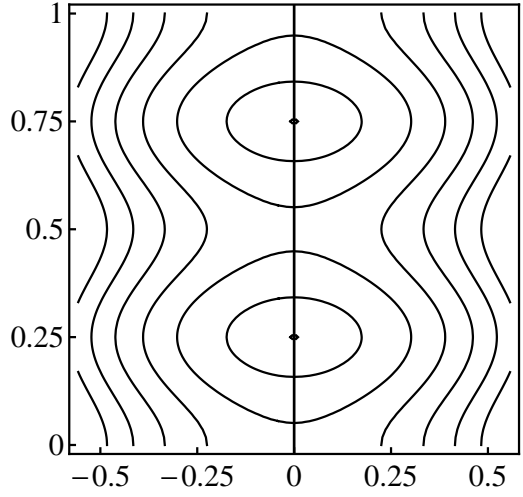}
\caption{Contour lines in the $xz$ plane of the ponderomotive potential for the standing Bessel wave ($M=0$). Owing to cylindrical symmetry the contour lines would be the same after an arbitrary rotation around the beam axis shown as the center line. The minima of the potential are marked by the dots.}\label{fig16}
\end{figure}

\begin{figure}
\centering
\includegraphics[scale=0.5]{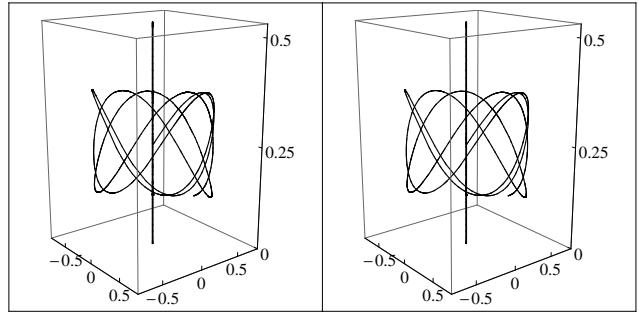}
\caption{Trajectory of a particle trapped in the vicinity of the potential minimum at $z=\pi/2k_\parallel$. Thick line is the beam axis. There is no noticeable difference between the exact solution (left) and the one obtained within the ponderomotive approximation (right).}\label{fig17}
\end{figure}

\begin{figure}
\centering
\includegraphics[scale=0.5]{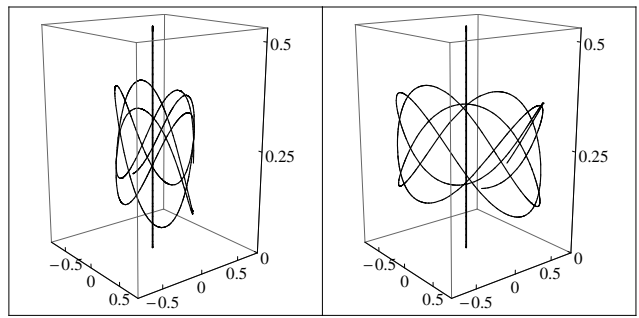}
\caption{Trajectories of a particle trapped in the vicinity of the potential minimum with the same parameters as in Fig.~\ref{fig17} but with the starting point moved away from the minimum. The trajectories are completely different since the adjustment of the initial velocity has not been made.}\label{fig18}
\end{figure}

\begin{figure}
\centering
\includegraphics[scale=0.5]{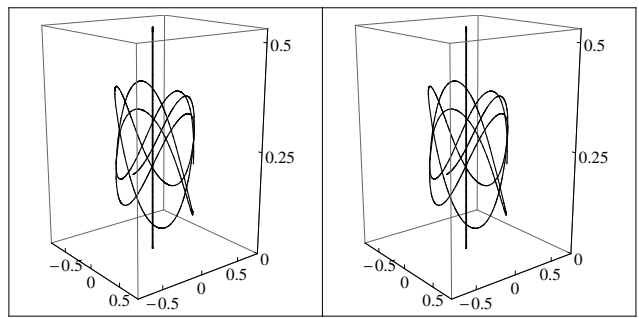}
\caption{Trajectories of a particle trapped in the vicinity of the potential minimum with the same parameters as in Fig.~\ref{fig17} but with the starting point moved away from the minimum. The agreement between the exact trajectory and the one in the ponderomotive approximations has been fully restored by adjusting the initial velocity.}\label{fig19}
\end{figure}

In all previous cases we considered trapping by running waves. In practical applications it may be easier to work with standing waves. It turns out that a standing Bessel wave can also fully trap charged particles. A standing Bessel wave is constructed by adding two Bessel beams that differ only in the sign of $k_\parallel$. The ponderomotive potential of this superposition is given by the formula
\begin{align}\label{psw}
V_p({\bm r})&=2V_0\big[2k_\parallel^2/k_\perp^2
\left(J_{M-1}^2(k_\perp\rho)+J_{M+1}^2(k_\perp\rho)\right)\nonumber\\
&+\left(1+\cos(2 k_\parallel z)\right)
\left(J_{M-1}^2(k_\perp\rho)
+J_{M+1}^2(k_\perp\rho)+2J_M^2(k_\perp\rho)\right)\big].
\end{align}
The ponderomotive potential is a periodic function of $z$ with the period equal to $\pi/k_\parallel$. The amplitude of these oscillations is exceedingly small. Owing to the large value of $k_\parallel^2/k_\perp^2$, the $z$-independent terms dominate. The case $M=0$ is exceptional since then the $z$-dependent term dominates near the beam axis. In Fig.~\ref{fig16} we show contour lines of the ponderomotive potential for $M=0$ cut by the $xz$ plane that exhibit the potential minima. The trajectories of a particle trapped in three dimensions near such a minimum are shown in Figs.~\ref{fig17}-\ref{fig19}. In Fig.~\ref{fig17} we have a fairly good agreement between the exact trajectory and its ponderomotive approximation because the starting point is very close to the potential minimum. In Fig.~\ref{fig18} the trajectories are completely different because the adjustment of the initial velocity has not been made. Finally, in Fig.~\ref{fig19} the adjustment has been made and the agreement has been fully restored.

\section{Outlook}

We have shown that Bessel beams might be used as a ``waveguide'' and also as a trap for charged particles. So far we only considered a single charged particle. The extension of our results to many particles is an open problem because it would require the inclusion of the Coulomb interaction between particles. The standing Bessel wave configuration offers a possibility of trapping many particles in successive minima of the ponderomotive potential. In this way one may construct an analog of the optical lattice for charged particles. The only condition for such a construction is that the Coulomb repulsion should not overcome the ponderomotive potential barrier. This means that the intensity of the beam must satisfy the inequality $I>I_c=4\pi m(c/\lambda)^3$. For the wave length $\lambda=6\cdot10^{-7}$m, this gives a rather high (but not totally unrealistic) value, $I_c=1.4\cdot 10^{15}{\rm W/m}^2$. When the intensity obeys the condition $I<2I_c$ then only one electron can be trapped at each potential minimum--- the Coulomb repulsion will push extra electrons away. One may significantly improve our standing wave trap by adding a constant magnetic field in the direction perpendicular to the beam axis. Such a magnetic field will also quantize the spin degree of freedom of electrons, transforming each electron into a qubit. All these conditions apply to electrons at rest. In order to hold electrons that undergo motion in a trap, we need higher intensities.

\section*{Acknowledgements}

We would like to thank Piotr Kucharski for useful comments. We acknowledge support from the Polish Ministry of Science and Higher Education under a grant for the years 2008-2010.


\begin{thebibliography}{2}
\bibitem{deh} H. Dehmelt, Experiment with an isolated subatomic particle at rest, Rev. Mod. Phys. {\bf 62}, 525 (1990).
\bibitem{paul} W. Paul, Electromagnetic traps for charged and neutral particles, Rev. Mod. Phys. {\bf 62}, 531 (1990).
\bibitem{bb}  I. Bialynicki-Birula, Particle beams guided by electromagnetic vortices: New solutions of the Lorentz, Schr\"odinger, Klein-Gordon, and Dirac equations, Phys. Rev. Lett. {\bf 93}, 020402 (2004).
\bibitem{bbch} I. Bialynicki-Birula, Z. Bialynicka-Birula, and B. Chmura, Trojan states of electrons guided by Bessel beams, Laser Phys. {\bf 15}, 1371 (2005).
\bibitem{br} I. Bialynicki-Birula and T. Rado\.zycki, Pinning and transport of cyclotron (Landau) orbits by electromagnetic vortices, Phys. Rev. A {\bf 73}, 052114 (2006).
\bibitem{mhj} V. H. Mellado, S. Hacyan, and R. J\'auregui, Trapping and acceleration of charged particles in Bessel beams, Laser and Particle Beams {\bf 24}, 1 (2006).
\bibitem{ger} D. Gerlich, Molecular ions and nanoparticles in RF and AC traps, Hyperfine Interactions {\bf 146/147}, 293 (2003).
\bibitem{pwf} I. Bialynicki-Birula, Progress in Optics, edited by E. Wolf (Elsevier, Amsterdam, 1996), Vol. 36; arXiv:quant-ph/0508202. This paper reviews the history of the RS vector.
\bibitem{trkal} V. Trkal, A note on the hydrodynamics of viscous fluids, Czech. J. of Phys. {\bf 44}, 97 (1994).
\bibitem{beams} I. Bialynicki-Birula and Z. Bialynicka-Birula, Beams of electromagnetic radiation carrying angular momentum: The Riemann-Silberstein vector and the classical-quantum correspondence, Opt. Comm. {\bf 263}, 342 (2006).
\bibitem{phloc} I. Bialynicki-Birula and Z. Bialynicka-Birula, Why photons cannot be sharply localized, Phys. Rev. A. {\bf 79}, 032112 (2009).
\bibitem{gm} E. V. Gaponov and M. E. Miller, On the Potential Wells for Charged Particles in a High-Frequency Electromagnetic Field, Zh. Eksp. Teor. Fiz. {\bf 34}, 242 (1958); JETP {\bf 7}, 68 (1958).
\bibitem{bh} H. A. H. Boot and R. B. R.-S. Harvie, Charged particles in a non-uniform radio-frequency field, Nature {\bf 180}, 1187 (1957).
\end{thebibliography}
\end{document}